\documentclass[11pt,twoside]{article}

\usepackage{galev06}
\usepackage{epsf}
\usepackage{psfig}
\usepackage{lscape}
\usepackage{graphicx}

\newcommand{\iso}{{\it ISO}}
\newcommand{\spitz}{{\it Spitzer}}

\newcommand{\irs}{{\it Spitzer/IRS}}
\newcommand{\isocam}{{\it ISO/CAM}}
\newcommand{\isosws}{{\it ISO/SWS}}
\newcommand{\M}[1]{M$\;$#1}
\newcommand{\ngc}[1]{NGC$\;$#1}
\newcommand{\hii}{H$\,${\sc ii}}
\newcommand{\snii}{SN$\;${\sc ii}}
\newcommand{\zsun}{\;Z_\odot}
\newcommand{\mic}{\;\mu m}
\newcommand{\E}[1]{\times10^{#1}}
\newcommand{\sms}[1]{{\mbox{{\scriptsize #1}}}}
\newcommand{\reffig}[1]{Fig.~\ref{#1}}
\newcounter{textlistctr}
\newcommand{\thetextlist}{, }
\newcommand{\textlist}[1]
            {\setcounter{textlistctr}{1}
             \renewcommand{\thetextlist}
             {{\it (\roman{textlistctr})}\stepcounter{textlistctr}}#1
              }

\begin{document}
\title{PAHs in Galaxies: their Properties and Evolution}
\author{Fr\'ed\'eric Galliano}
\affil{Observational Cosmology Lab., Code 665, 
       NASA Goddard Space Flight Center, 
       Greenbelt MD 20771, USA}

\begin{abstract}
I summarize the results of two recent studies, based on \iso\ and \spitz\
mid-IR spectra of galaxies and Galactic regions, aimed at understanding the 
origins of the variations of the aromatic features among and inside galaxies.
I show that the ratios between the most intense bands ($6.2, 7.7, 8.6$ and $11.3\mic$)
are principally sensitive
to the charge of the molecules, and therefore represent a powerful diagnostic
tool of the physical conditions inside the region where the emission is 
originating.
Then, I show that the weakness of the aromatic bands, in low-metallicity 
environments, is a consequence of the delayed injection of their carriers, 
the Polycyclic Aromatic Hydrocarbons (PAHs), into the interstellar medium (ISM)
of galaxies.
Indeed, PAHs are believed to form in the envelopes of post-AGB stars,
several hundreds of million years after the beginning of the star formation,
when the system is already chemically evolved.
\end{abstract}

\section{Introduction}

The mid-infrared (mid-IR; $\lambda\simeq3-30\mic$) range of the electromagnetic
spectrum is 
dominated, in a wide variety of objects, by broad emission features, generally 
attributed to the vibrational modes of Polycyclic Aromatic Hydrocarbons 
\citep[][for a review]{peeters04b}.
These PAHs are large planar molecules, made of 50 to 1000 carbon atoms, 
stochastically heated by single photon events.
In our Galaxy, they are responsible for $\sim 15\;\%$ of the reprocessing of
the stellar light by dust, and contain $\sim 15-20\;\%$ of the depleted carbon 
\citep[][with solar abundance constraints]{zubko04}.
In addition, they are major contributors to the photoelectric heating of 
the gas \citep[{e.g.}][]{hollenbach97}.

From an observational point of view, the complexity of the aromatic feature
spectrum gives potentially access to 
the physical conditions and the history of the region where it is originating.
In this review, I will focus essentially on two aspects of this 
spectrum:\textlist{\thetextlist~the ratio between the most intense bands
           ($6.2, 7.7, 8.6$ and $11.3\mic$) and
         \thetextlist~the intensity of the bands relative to the underlying
           continuum.}This 
work is based on a sample of nearby galaxies and Galactic regions, observed
with \isocam, \isosws\ and \irs.

\section{Tracing the Physical Conditions with the Aromatic Features}

The detailed study of the aromatic features requires an accurate measure of the
intensity of the various bands, on a given mid-IR spectrum.
However, the intrinsic width of the bands ($\Delta\lambda\simeq 1\mic$), their
overlap, their dilution by the continuum, and the presence of the amorphous
silicate feature at $9.7\mic$, makes their absolute measure uncertain.
The main difficulty is to separate the continuum, from the wings of the aromatic
bands.
To remain conservative, we systematically used two separate methods to 
decompose each one of our 
spectra:\textlist{\thetextlist~a method that underestimates the contribution
          of the wings of the bands 
          \citep[inspired by the one described by][]{vermeij02} and
        \thetextlist~a method that overestimates these wings, by letting
          the width of the bands vary freely 
          \citep[inspired by the one described by][]{laurent00}.}The
physical results obtained with the two methods are similar. 
The numerical values of the band intensities change from one method to 
the other, but the correlations between the various band ratios, and their 
trends with the physical properties are identical.
Therefore, to be concise, I will present here only the results obtained with
the second method and refer the reader to the upcoming complete study 
\citep{galliano06}.

\begin{figure}[htbp]
  \centering
  \includegraphics[width=\textwidth]{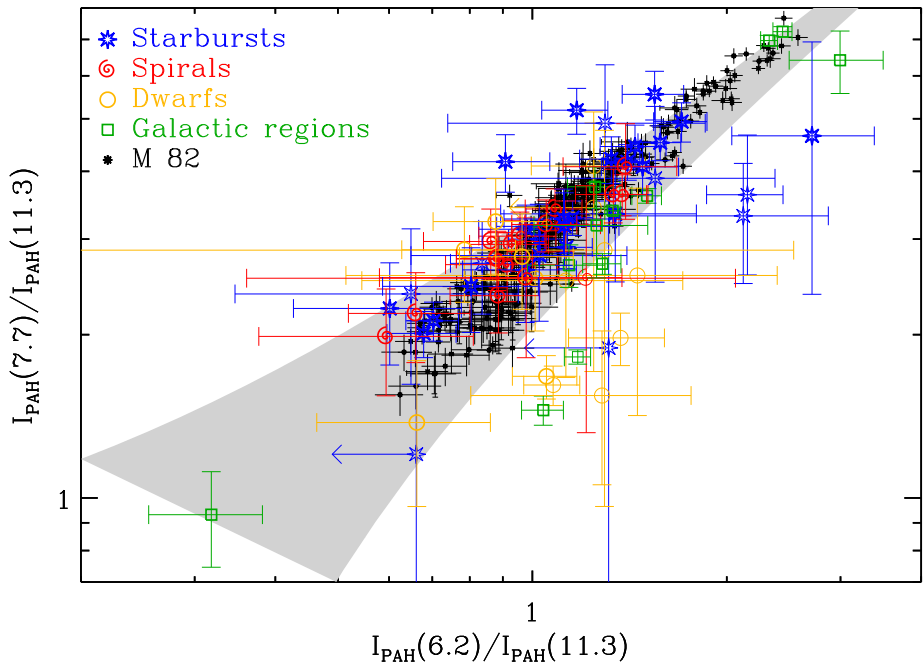}
  \caption{Correlation between two band ratios.
           The colored symbols, with the error bars, are integrated spectra
           on various sources.
           The black symbols are measurements on various regions inside \M{82}.
           The grey area is the $\pm1\sigma$ linear correlation to the cloud
           of points.
           The faint symbols are the points which have been excluded, because
           of either a low feature-to-continuum ratio, or a low signal-to-noise
           ratio.}
  \label{fig:correlation}
\end{figure}
We have studied the intensities of the $6.2, 7.7, 8.6$ and $11.3\mic$ bands
on the individual spectra of quiescent spirals, starburst and dwarf galaxies,
and Galactic \hii\ regions, planetary nebulae, and photodissociation regions 
(PDRs).
Our most important result is that the ratios between the $6.2, 7.7$ and 
$8.6\mic$ bands do not significantly vary throughout our sample, while the 
ratios between these three bands and the $11.3\mic$ band spread over one order
of magnitude (e.g.\ \reffig{fig:correlation}).
Laboratory studies on PAH analogs conclude that these variations are likely
controlled by the fraction of ionized PAHs \citep[see][]{allamandola89}.
However, they could also be explained by a depletion of the smallest
molecules.
Using a stochastic heating model, and realistic optical properties, we were able
to show that this explanation was not consistent with our data.
Indeed, a variation of the $6.2/11.3$ ratio, by depleting the smallest PAHs, 
would result in a significant variation of the $6.2/7.7$ and $6.2/8.6$ ratios,
which we found to be remarkably constant.
Similarly, a variation of the $6.2/11.3$ ratio due to the extinction by
the $9.7\mic$ feature can be ruled out, since it would also affect the 
$6.2/8.6$ ratio.
\begin{figure}[htbp]
  \centering
  \includegraphics[width=\textwidth]{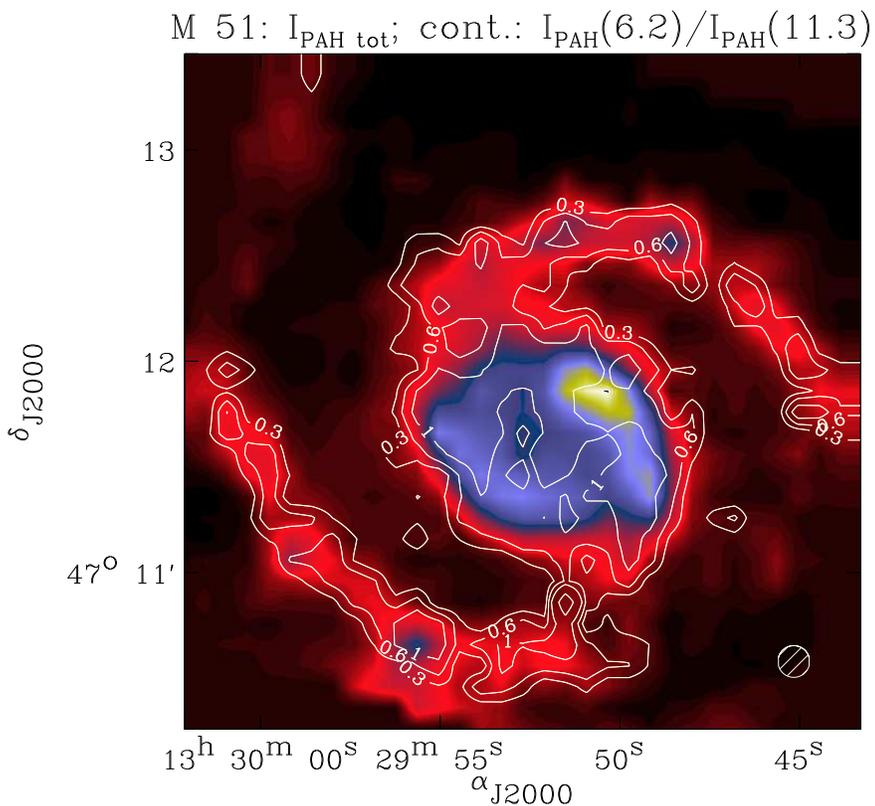}
  \caption{The color image is the spatial distribution of
           the PAH intensity inside \M{51}, and the contours represent the ratio
           between the $6.2$ and $11.3\mic$ bands \citep{galliano06}.}
  \label{fig:correlation2}
\end{figure}

\begin{figure}[htbp]
  \includegraphics[width=\textwidth]{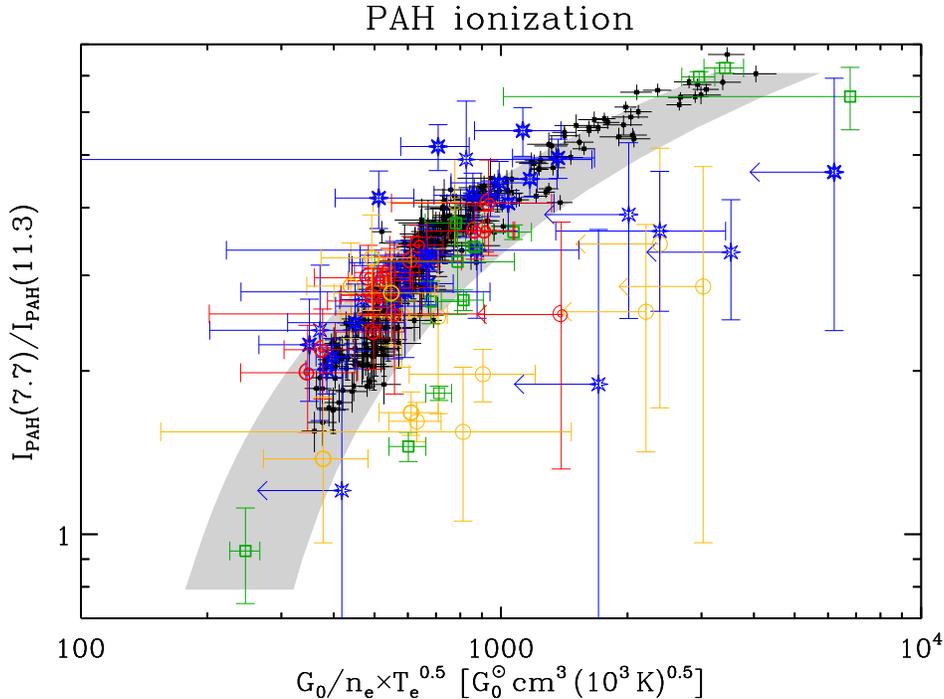}
  \caption{Relation between a PAH band ratio and the physical conditions
           \citep{galliano06}.
           The solar neighborhood UV density is
           $G_0^\odot = 1.6\E{-6}\;\rm W\,m^{-2}$;
           $n_e$ is the electronic density in $\rm cm^{-3}$; and
           $T_e$ is the electronic temperature in K.}
  \label{fig:correlation3}
\end{figure}

We were able to show that the properties of the PAHs, inside galaxies
of different types, are remarkably universal, and that the variations of the
band ratios are essentially due to the fraction of ionized PAHs.
Therefore, this band ratio can be used to probe the physical conditions into
the regions where the emission is originating.
We used well-studied Galactic regions, like the Orion bar, \ngc{7027}, and 
\ngc{2023}, to give an empirical correspondance between the value of the
band ratio, and the ratio $G_0/n_e\times\sqrt{T_e}$ 
\citep[\reffig{fig:correlation3}; see also][]{bregman05}.
The quantity $G_0/n_e\times\sqrt{T_e}$, involving the UV field density 
$G_0$, the electronic density $n_e$, and the electronic temperature $T_e$, is a physical parameter quantifying the balance between ionization and recombination.
It is one of the most important parameters of PDR models.
At the scale of a galaxy, when the clouds are not resolved, this quantity
can potentially be used to constrain the filling factor and density of the PDRs.

\section{The Delayed Injection of PAHs by AGB Stars}

Pioneer ground-based observations \citep{roche91}, \iso\ 
\citep{thuan99,galliano04,madden06} and \spitz\ studies 
\citep{engelbracht05,wu06,ohalloran06} have shown that PAHs were 
underabundant in low-metallicity environments.
Massive destruction of the PAHs by the hard radiation field, amplified by the lower dust screening, was proposed by \citet{galliano03,galliano05} and 
\citet{madden06} to explain this property.
Conversely, \citet{ohalloran06} proposed that PAHs, in dwarf galaxies, 
could be significantly depleted by the numerous \snii\ blast waves.
Finally, \citet{dwek05} suggested that the paucity of PAHs, in chemically young
systems, is a consequence of their late injection by long-lived low-mass stars.

In order to address this problem, we modeled the UV-to-millimetre spectral energy distributions (SEDs) of 35 nearby galaxies, with metallicities ranging
from $1/50$ to $3\zsun$.
We studied the evolution of the dust-to-gas mass ratios of the PAHs 
($Z_\sms{PAH}$) and of the far-IR emitting dust ($Z_\sms{FIR dust}$), as 
a function of the metallicity of the ISM ($Z_\sms{ISM}$).

\begin{figure}[htbp]
  \includegraphics[width=\textwidth]{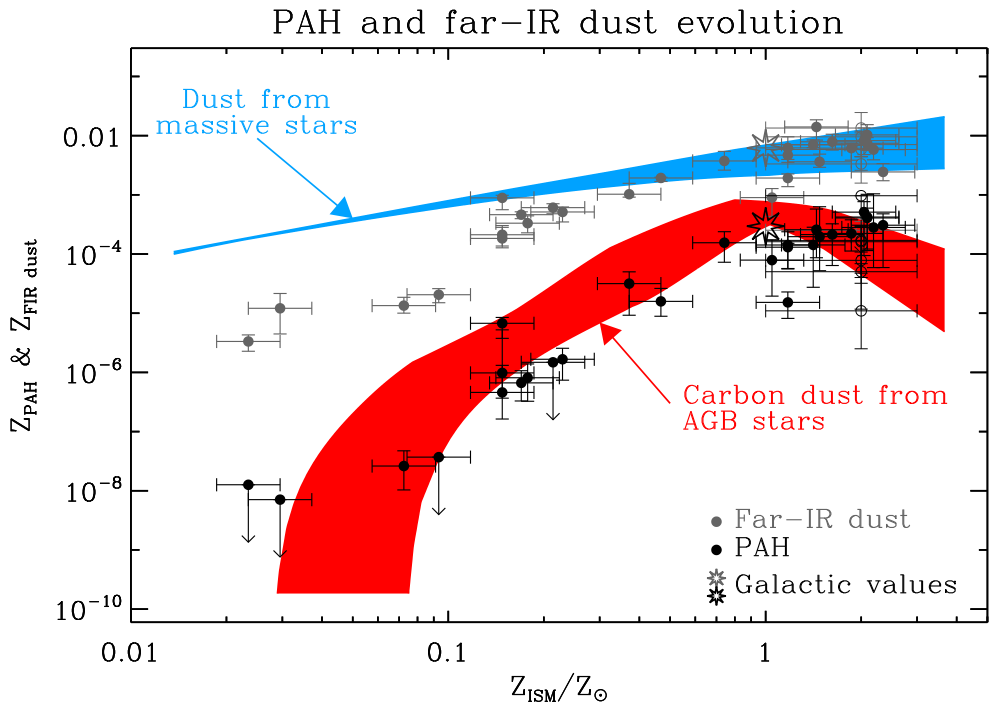}
  \caption{Evolution of the PAH- and FIR-dust-to-gas mass ratios derived 
           from the SED modelling, compared to the production rate of
           carbon dust by AGB stars, and to the one of total dust by
           massive stars \citep{galliano06b}.}
  \label{fig:pahvsZ}
\end{figure}
We investigated the resulting trends, by comparing our findings to the outputs
of a dust evolution model \citep{galliano06b}.
\reffig{fig:pahvsZ} shows this comparison.
We showed that the PAH content is remarkably correlated with the rate of carbon
dust production by AGB stars.
Indeed, PAHs are believed to form in the envelopes of AGB stars.
These AGB stars, which have a lifetime of several hundred million years, inject 
their material into the ISM, when the system is already evolved.
This behavior corroborates naturally the paucity of PAHs, in low-metallicity
environments.
The FIR-dust content follows very well the dust production rate by massive stars
down to $0.1\zsun$.
The discrepancy, that we observe below $0.1\zsun$, could either be due to
an overlooked, significant amount of cold dust in dwarf galaxies 
\citep[{e.g.}][]{galliano03,galliano05}, or a discontinous star formation 
history of the youngest objects \citep[{e.g.}][]{legrand00}.

\acknowledgements
I thank Rachel Dudik for her comments on this paper.
This work has been performed while I held a NRC Research Associateship Award,
and subsequently a NASA Postdoctoral Fellowship at NASA GSFC.

\bibliographystyle{galev06}
\bibliography{/Users/Fred/Documents/Astro/TeXStyle/references}

\end{document}